\documentclass[a4paper]{llncs}
\usepackage{amsmath, amsfonts, amssymb, comment}
\usepackage{geometry}

\newtheorem{thm}{Theorem}
\newtheorem{cor}[thm]{Corollary}
\newtheorem{prop}[thm]{Proposition}
\newtheorem{lem}[thm]{Lemma}

\newtheorem{defi}[thm]{Definition}

\newtheorem{ex}[thm]{Example}

\newcommand{\rs}{\mathrm{rs}}
\newcommand{\F}{\mathbb{F}}
\newcommand{\G}{\mathcal{G}(k,n)}

\begin{document}

\title{A Complete Characterization of \\
  Irreducible Cyclic Orbit
  Codes\thanks{Research partially supported by Swiss National Science
    Foundation Project no. 126948}} %
\author{Anna-Lena Trautmann \and Joachim Rosenthal}
\institute{Institute of Mathematics\\
  University of Zurich, Switzerland\\
  \email{www.math.uzh.ch/aa}}
\maketitle


\begin{abstract}
  Constant dimension codes are subsets of the finite Grassmann
  variety. The study of constant dimension codes with good distances
  have been central in random linear network coding theory.

  Orbit codes represent a subclass of constant dimension codes. They are characterized that the elements of the code can
  be viewed as the orbit under a group action. 

  The paper gives a complete characterization of orbit codes that are
  generated by an irreducible cyclic group, i.e. an irreducible group
  having one generator. We show how some of the basic properties of
  these codes, the cardinality and the minimum distance, can be derived using the isomorphism of the vector space and the extension field.
\end{abstract}

\section{Introduction}

In network coding one is looking at the transmission of information
through a directed graph with possibly several senders and several
receivers \cite{ah00}. One can increase the throughput by linearly
combining the information vectors at intermediate nodes of the
network.  If the underlying topology of the network is unknown we
speak about \textit{random linear network coding}. Since linear spaces
are invariant under linear combinations, they are what is needed as
codewords \cite{ko08}. It is helpful (e.g. for decoding) to constrain
oneself to subspaces of a fixed dimension, in which case we talk about
\emph{constant dimension codes}.

The set of all $k$-dimensional subspaces of a vector space $V$ 
is often referred to as the Grassmann variety (or simply Grassmannian) 
and denoted by 
$\mathcal{G}(k,V)$. {\em Constant dimension codes} are 
subsets of $\mathcal{G}(k,\F_{q}^n)$, where $\F_{q}$ is some finite field.

The general linear group $GL(V)$ consisting of all invertible
transformations acts naturally on the Grassmannian $\mathcal{G}(k,V)$.
If $\mathfrak{G}\leq GL(\F_{q}^n)=GL_n$ is a subgroup 
then one has an induced action of  $\mathfrak{G}$ on the 
finite Grassmannian $\mathcal{G}(k,\F_{q}^n)$. Orbits under 
the $\mathfrak{G}$-action are called \textit{orbit
  codes} \cite{tr10p}. The set of orbit codes comes with nice
algebraic properties. E.g. for the computation of the distance 
of an orbit code it is enough to compute the distance between
the base point and any of its orbit elements. This is in analogy to
linear block codes where the distance can also be computed from the
weights of the non-zero code words.

Orbit codes can be classified according to the groups used to
construct the orbit. In this work we characterize orbit codes
generated by irreducible cyclic subgroups of the general linear group.

The paper is structured as follows: The second section gives some
preliminaries, first of random network coding and orbit codes. Then
some facts on irreducible polynomials are stated and the
representation of finite vector spaces as Galois extension fields is
explained in \ref{galois}. In part \ref{img} we introduce irreducible
matrix groups and give some properties, with a focus on the cyclic
ones.  The main body of the paper is Section \ref{icoc}, where we study
the behavior of orbit codes generated by these groups and compute the
cardinality and minimum distances of them. We begin by characterizing
primitive orbit codes and then study the non-primitive irreducible
ones.  Finally we give a conclusion and an outlook in Section \ref{conclusion}.

\section{Preliminaries}
\subsection{Random Network Codes}\label{oc}

Let $\mathbb{F}_q$ be the finite field with $q$ elements (where
$q=p^r$ and $p$ prime). For simplicity we will denote the Grassmannian $\mathcal{G}(k,\F_{q}^n)$ by $\G$. The general linear group of
dimension $n$, $GL_n$, is the set of all invertible $n\times
n$-matrices with entries in $\F_{q}$. Moreover, the set of all $k\times n$-matrices over $\F_q$ is denoted by $Mat_{k\times n}$.

Let $U\in Mat_{k\times n}$ be a matrix of rank $k$ and
\[
\mathcal{U}=\rs (U):= \text{row space}(U)\in \G.
\]
One can notice that the row space is invariant under $GL_k$-multiplication from
the left, i.e. for any $T\in GL_k$
\[
\mathcal{U}=\rs(U)= \rs(T U).
\]
Thus, there are several matrices that represent a given subspace. A
unique representative of these matrices is the one in reduced row
echelon form.
Any $k\times n$-matrix can be transformed into reduced row echelon
form by a $T\in GL_k$.

The \emph{subspace distance}  is a metric on $\G$
given by
\begin{align*}
  d_S(\mathcal{U},\mathcal{V}) =& 2(k - \dim(\mathcal{U}\cap
  \mathcal{V}))\\
  =& 2\cdot \mathrm{rank}\left[ \begin{array}{c} U \\ V \end{array}
  \right] - 2k
\end{align*}
for any $\mathcal{U},\mathcal{V} \in \G$ and some
respective matrix representations $U$ and $V$. It is a suitable
distance for coding over the erasure channel \cite{ko08,si08a}.

A \textit{constant dimension code} $\mathcal{C}$ is simply a subset of the
Grassmannian $\G$. The minimum distance is defined
in the usual way. A code $\mathcal{C}\subset \G$ with minimum
distance $d_S(\mathcal{C})$ is called an $[n, d_S(\mathcal{C}), |\mathcal{C}|, k]$-code.

In the case that $k$ divides $n$ one can construct \textit{spread
  codes} \cite{ma08p}, i.e. optimal codes with minimum distance $2k$.
These codes are optimal because they reach the Singleton-like bound,
which means they have $\frac{q^{n}-1}{q^{k}-1}$ elements.

Given $U\in Mat_{k\times n}$ of rank $k$,
$\mathcal{U}\in \G$ its row space and $A\in GL_n$, we
define
\[
\mathcal{U} A:=\rs(U A).
\]

Let $U,V\in Mat_{k\times n}$ be matrices such that
$\rs(U)=\rs(V)$. Then one readily verifies that $\rs(U A)=\rs(V A)$
for any $A\in GL_n$.

This multiplication with matrices from $GL_n$ actually defines a
group operation from the right on the Grassmannian:

\[
\begin{array}{ccc}
  \G\times GL_n& \longrightarrow &
  \G \\ 
  (\mathcal{U},A) & \longmapsto & \mathcal{U} A
\end{array}
\]
Let $\mathcal{U}\in \G$ be fixed and $\mathfrak{G}$ a
subgroup of $GL_n$. Then
\[
\mathcal{C}= \{\mathcal{U} A \mid A \in \mathfrak{G}\}
\]
is called an \emph{orbit code} \cite{tr10p}. Since
\[
\G \cong GL_n/Stab(\mathcal{U}),
\]
where $Stab(\mathcal{U}):=\{A\in GL_n | \mathcal{U} A=\mathcal{U}\}$, 
it is possible that different subgroups generate the same orbit code.
An orbit code is called \emph{cyclic} if it can be defined by a cyclic
subgroup $\mathfrak{G} \leq GL_n$.

\subsection{Irreducible Polynomials and Extension
  Fields}\label{galois}

Let us state some known facts on irreducible polynomials over finite
fields (cf. \cite{li86} Lemmas 3.4 - 3.6):

\begin{lem}
  Let $p(x)$ be an irreducible polynomial over $\mathbb{F}_{q}$ of
  degree $n$, $p(0)\neq 0$ and $\alpha$ a root of it. Define the order
  of a polynomial $p(x) \in \F_{q}[x]$ with $p(0)\neq 0$ as the
  smallest integer $e$ for which $p(x)$ divides $x^e-1$.  Then
  \begin{enumerate}
  \item the order of $p(x)$ is equal to the order of $\alpha$ in
    $\F_{q^{n}}\backslash \{0\}$.
  \item the order of $p(x)$ divides $q^n-1$.
  \item $p(x)$ divides $x^c-1$ iff the order of $p(x)$ divides $c$
    (where $c\in \mathbb{N}$).
  \end{enumerate}

\end{lem}

There is an isomorphism between the vector space $\mathbb{F}_q^n$ and
the Galois extension field $\mathbb{F}_{q^n} \cong
\mathbb{F}_q[\alpha]$, for $\alpha$ a root of an irreducible
polynomial $p(x)$ of degree $n$ over $\mathbb{F}_q$. If in addition
$p(x)$ is primitive, then
\[\mathbb{F}_q[\alpha]\backslash\{0\}=\langle \alpha \rangle =
\{\alpha^i | i=0,...,q^n-2\}\] i.e.  $\alpha$ generates
multiplicatively the group of invertible elements of the extension
field.

\begin{lem}\label{vsprim}
  If $k|n, c:=\frac{q^n-1}{q^k-1}$ and $\alpha$ a primitive element of
  $\mathbb{F}_{q^n}$, then the vector space generated by
  $1,\alpha^{c},..., \alpha^{(k-1)c}$ is equal to $\{\alpha^{ic} |
  i=0,...,q^{k}-2\}\cup\{0\}=\F_{q^k}$.
\end{lem}
\begin{proof}
  Since $k|n$ it holds that $c\in \mathbb{N}$.  Moreover, it holds that
  $(\alpha^{c})^{q^{k}-1}= \alpha^{q^{n}-1}= 1$ and
  $(\alpha^c)^{q^{k}-2}= \alpha^{-c}\neq 1$, hence the order of
  $\alpha^{c}$ is $q^{k}-1$.  It is well-known that if $k$ divides $n$
  the field $\mathbb{F}_{q^n}$ has exactly one subfield
  $\mathbb{F}_{q^k}$. Thus the group generated by $\alpha^{c}$ has to
  be $\mathbb{F}_{q^k}\backslash \{0\}$, which again is isomorphic to
  $\mathbb{F}_q^{k}$ as a vector space.  \qed \end{proof}



\subsection{Irreducible Matrix Groups}\label{img}

\begin{defi}
  \begin{enumerate}
  \item A matrix $A\in GL_n$ is called \textsl{irreducible} if
    $\F_q^n$ contains no non-trivial $A$-invariant subspace, otherwise it is
    called \textsl{reducible}.
  \item A subgroup $\mathfrak{G}\subseteq GL_n$ is called
    \textsl{irreducible} if $\F_q^n$ contains no
    $\mathfrak{G}$-invariant subspace, otherwise it is called
    \textsl{reducible}.
   \item  An orbit code $\mathcal{C} \subseteq \G$ is called 
    \textsl{irreducible} if $\mathcal{C}$ can be viewed as the orbit 
    under the group action of an irreducible group.
  \end{enumerate}
\end{defi}

A cyclic group is irreducible if and only if its generator matrix is
irreducible. Moreover, an invertible matrix is irreducible if and only
if its characteristic polynomial is irreducible.

\begin{ex} 
  Over $\mathbb{F}_2$ the only irreducible polynomial of degree $2$ is
  $p(x)=x^2+x+1$. The irreducible matrices in $GL_2$ must have trace
  and determinant equal to $1$ and hence are
  \[\left(\begin{array}{cc} 0 & 1 \\ 1 & 1 \end{array}\right)
  \textnormal{ and } \left(\begin{array}{cc} 1 & 1 \\ 1 & 0
    \end{array}\right).\]
\end{ex}

We can say even more about irreducible matrices
with the same characteristic polynomial.
For this, note that the definition of an irreducible matrix $G$
implies the existence of a so called {\em cyclic vector} $v\in \F_{q}^n$
having the property that
$$
\{ v,vG,vG^2,\ldots,vG^{n-1}\}
$$
forms a basis of $\F_{q}^n$. Let $S\in GL_n$ be the basis transformation
which transforms the matrix $G$ into this new basis. Then it follows that
$$
SGS^{-1}=
\left( 
  \begin{array}{ccccc}
    0 & 1 & 0 &\ldots  &0\\
    0 & 0 & 1 &\ldots  &0\\
 \vdots&\vdots&\vdots&\ddots&\vdots\\
   0 & 0 & 0 &\ldots  &1\\
   -c_0 & -c_{1} & & \ldots  &-c_{n-1}
  \end{array}
\right).
$$
The matrix appearing on the right is said to be
in {\em companion form}. By convention we will use row
vectors and accordingly companion matrices where the coefficients of
the corresponding polynomials are in the last row (instead of the last
column).

One readily verifies that 
$$
p(x):=x^n+c_{n-1}x^{n-1}+\cdots+c_1x+c_0
$$
is the characteristic polynomial of both $G$ and $SGS^{-1}$.  It
follows that every irreducible matrix in $GL_n$ is similar to the
companion matrix of its characteristic polynomial. Hence all
irreducible matrices with the same characteristic polynomial are
similar.

Furthermore, we can say something about the order of a matrix when viewed as
an element of the finite group $GL_n$. For this assume that
$G\in GL_n$ is an invertible matrix having $p(x)$ as
characteristic polynomial. Then one readily verifies that the order of
$G$ is equal to the order of $p(x)$.  Hence $G$ is a primitive element
of $GL_n$ if and only if its characteristic polynomial is primitive.

The next fact is a well-known group theoretic result:

\begin{lem}(cf. \cite{li86} Theorem 1.15.)  In a finite cyclic group
  $\mathfrak{G}=\langle G \rangle$ of order $m$, the element $G^l$
  generates a subgroup of order $\frac{m}{\gcd(l,m)}$. Hence each
  element $G^l$ with $\gcd(l,m)=1$ is a generator of $\mathfrak{G}$.
\end{lem}

\begin{lem}(cf. \cite{ma11} Theorem 7)
   All irreducible cyclic groups generated by matrices with a
   characteristic polynomial of the same order are conjugate to each
   other.
\end{lem}

\begin{ex}
  Over $\mathbb{F}_2$ the irreducible polynomials of degree $4$ are
  $p_1(x)=x^4+x+1$, $p_2(x)=x^4+x^3+1$ and $p_3(x)=x^4+x^3+x^2+x+1$, where
  $ord(p_1)=ord(p_2)=15$ and $ord(p_3)=5$. Let $P_1,P_2,P_3$ be the
  respective companion
  matrices: 
  One verifies that $\langle P_1\rangle$ and $\langle P_2\rangle$ are
  conjugate to each other
  but $\langle P_3 \rangle$ is not conjugate to them.

\end{ex}

One can describe the action of an irreducible matrix group via the
Galois extension field isomorphism.
\begin{thm}\label{compmult}
  Let $p(x)$ be an irreducible polynomial over $\mathbb{F}_{q}$ of
  degree $n$ and $P$ its companion matrix.  Furthermore let $\alpha
  \in \mathbb{F}_{q^n} $ be a root of $p(x)$ and $\phi$ be the
  canonical homomorphism
  \begin{align*}
    \phi: \mathbb{F}_q^n &\longrightarrow \mathbb{F}_{q^n} \\
    (v_1,\dots,v_n) &\longmapsto \sum_{i=1}^n v_i \alpha^{i-1} .
  \end{align*}
  Then the following diagram commutes (for $v \in \mathbb{F}_q^n $):
  \[\begin{array}{rcl}
    v &\overset{ P}{\longrightarrow} &vP\\
    \phi \downarrow & &\downarrow \phi \\
    v'    &\underset{ \alpha}{\longrightarrow} & v'\alpha
  \end{array}\]
\end{thm}

If $P$ is a companion matrix of a primitive polynomial the group
generated by $P$ is also known as a \textit{Singer group}. This
notation is used e.g. by Kohnert et al. in their network code
construction (see \cite{el10}, \cite{ko08p}). Elsewhere $P$ is called \textit{Singer
  cycle} or \textit{cyclic projectivity}  (e.g. in \cite{hi98}).

\section{Irreducible Cyclic Orbit Codes}\label{icoc}

The irreducible cyclic subgroups of $GL_n$ are exactly the groups generated
by the companion matrices of the irreducible polynomials of degree $n$
and their conjugates. Moreover, all groups generated by companion matrices of irreducible
polynomials of the same order are conjugate.

It is sufficient to characterize the orbits of cyclic groups generated
by companion matrices of irreducible polynomials of degree $n$.  The
following theorem shows that the results are then carried over to any
irreducible cyclic orbit code via the choice of a starting point of
the orbit.

\begin{thm}
  Let $G$ be an irreducible matrix, $\mathfrak{G} =\langle G \rangle$
  and $\mathfrak{H} =\langle S^{-1}GS \rangle$ for an $S\in GL_n$.
  Moreover, let $\mathcal{U} \in \G$ and $\mathcal{V} :=
  \mathcal{U}S$. Then the orbit codes
  \[\mathcal{C}:=\{\mathcal{U}A | A \in \mathfrak{G} \} \textnormal{
    and } \mathcal{C'}:=\{\mathcal{V}B | B \in \mathfrak{H} \}\] have
  the same cardinality and minimum distance.
\end{thm}
\begin{proof}
  Trivially the cardinality of both codes is the same. It remains to
  be shown that the same holds for the minimum distance.

  The following diagram commutes:
  \[\begin{array}{rcl}
    \mathcal{U} &\overset{G}{\longrightarrow} & \mathcal{U}G\\
    S\downarrow & &\downarrow S \\
    \mathcal{V}  &\underset{S^{-1}GS}{\longrightarrow} &  \mathcal{U}GS
  \end{array}\]
  Since the subspace distance is invariant under $GL_n$-action and
  $(S^{-1}GS)^i=S^{-1}G^iS$ it holds that
  \[d_S(\mathcal{U}, \mathcal{U}G^i)=d_S(\mathcal{V},\mathcal{U}G^i
  S)\] hence the minimum distances of the codes defined by
  $\mathfrak{G}$ and by $\mathfrak{H}$ are equal.  \qed \end{proof}

\subsection{Primitive Generator}\label{prim}

Let $\alpha$ be a primitive element of $\F_{q^n}$ and assume $k|n$ and
$c:=\frac{q^n-1}{q^k-1}$. Consider once more the $\F_q$-subspace
$\F_{q^k}=\{\alpha^{ic} | i=0,...,q^{k}-2\}\cup\{0\}$.

\begin{lem}
  For every $\beta\in\F_{q^n}$ the set
$$
\beta\cdot\F_{q^k}=\{\beta\alpha^{ic} | i=0,...,q^{k}-2\}\cup\{0\}
$$
defines an $\F_q$-subspace of dimension $k$.
\end{lem}
\begin{proof}
\begin{align*}
\varphi_\beta:\ \F_{q^n} & \longrightarrow \F_{q^n}\\ 
u &\longmapsto \beta u
\end{align*}
is an $\F_q$-linear isomorphism,
$\varphi_\beta(\F_{q^k})=\beta\cdot\F_{q^k}$ is hence an $\F_q$-linear
subspace of dimension $k$.\qed
\end{proof}

\begin{thm}
 The set
$$
\mathcal{S}=\left\{ \alpha^i\cdot\F_{q^k}\mid i=0,\ldots,c-1\right\}
$$
defines a spread code. 
\end{thm}
\begin{proof}
  By a simple counting argument it is enough to show that the subspace
  $\alpha^i\cdot\F_{q^k}$ and $\alpha^j\cdot\F_{q^k}$ have only
  trivial intersection whenever $0\leq i<j\leq c-1$.  For this assume
  that there are field elements $c_i,c_j\in \F_{q^k}$, such that
$$
v=\alpha^i c_i=\alpha^j c_j\in \alpha^i\cdot\F_{q^k}\cap
\alpha^j\cdot\F_{q^k}.
$$
If $v\neq 0$ then $\alpha^{i-j}=c_j c_i^{-1}\in \F_{q^k}.$ But this
means $i-j\equiv 0 \mod c$ and $ \alpha^i\cdot\F_{q^k}=
\alpha^j\cdot\F_{q^k}$.  It follows that $\mathcal{S}$ is a spread.
\qed
\end{proof}

We now translate this result into a matrix setting.
For this let $\phi$  denote the canonical homomorphism as
defined in Theorem \ref{compmult}.

\begin{cor}\label{spread}
  Assume $k|n$.  Then there is a subspace 
 $\mathcal{U} \in \G$ such that the cyclic orbit code obtained by
  the  group action of a 
 a primitive companion matrix is a code with
  minimum distance $2k$ and cardinality $\frac{q^n-1}{q^k-1}$. Hence
  this  irreducible cyclic orbit code is a spread code.
\end{cor}
\begin{proof}
  In the previous theorem represent $\F_{q^k}\subset \F_{q^n}$
  as the rowspace of a $k\times n$ matrix over $\F_q$ and using
  the same basis over $\F_q$ represent the primitive $\alpha$ with
  a primitive companion matrix $P$. The orbit code defined in this
way has then all the desired properties.
  \qed \end{proof}

\begin{ex}
  Over the binary field let $p(x):=x^6+x+1$ be primitive, $\alpha$ a root
  of $p(x)$ and $P$ its companion matrix.
  \begin{enumerate}
  \item For the 3-dimensional spread compute $c=\frac{63}{7}=9$ and
    construct a basis for the starting point of the orbit:
    \begin{align*}
      u_1&=\phi^{-1} (\alpha^0)=\phi^{-1} (1)=(100000)\\
      u_2&=\phi^{-1} (\alpha^c)=\phi^{-1}(\alpha^9)=\phi^{-1}(\alpha^4+\alpha^3)=(000110)\\
      u_3&=\phi^{-1} (\alpha^{2c})=\phi^{-1}(\alpha^{18})=\phi^{-1}(
      \alpha^3+\alpha^2+\alpha+1)= (111100)
    \end{align*}
    The starting point is
    \[ \mathcal{U}=\rs\left[\begin{array}{cccccc} 1&0&0&0&0&0\\
        0&0&0&1&1&0\\ 1&1&1&1&0&0 \end{array}\right] =
    \rs\left[\begin{array}{cccccc} 1&0&0&0&0&0\\ 0&1&1&0&1&0 \\
        0&0&0&1&1&0 \end{array}\right] \] and the orbit of the group
    generated by $P$ on $\mathcal{U}$ is a spread code.
  \item For the 2-dimensional spread compute $c=\frac{63}{3}=21$ and
    construct the starting point
    \begin{align*}
      u_1&=\phi^{-1} (\alpha^0)=\phi^{-1}(1)=(100000)\\
      u_2&=\phi^{-1} (\alpha^c)=\phi^{-1}(\alpha^{21})=\phi^{-1}(\alpha^2+\alpha+1)=
      (111000)
    \end{align*}
    The starting point is
    \[\mathcal{U}=\rs\left[\begin{array}{cccccc} 1&0&0&0&0&0\\
        1&1&1&0&0&0 \end{array}\right]=\rs\left[\begin{array}{cccccc}
        1&0&0&0&0&0\\ 0&1&1&0&0&0 \end{array}\right]\]
    and the orbit of the group generated by $P$ is a spread code.
  \end{enumerate}
\end{ex}

The following fact has been formulated by Kohnert and Kurz in
\cite{ko08p}:

\begin{lem}\label{intone}
  Over the binary field let $p(x)$ be a primitive polynomial and
  $\alpha$ a root of it. Assume $k>1$ and
  $\mathcal{U}=\{0,u_1,\dots,u_{2^k-1}\} \in \G$ such that
  \[\phi(u_i)=\alpha^{b_i} \hspace{0.7cm} \forall i=1,\dots,2^k-1\]
  and the set
  \[\{b_{m}- b_{l} \mod 2^n-1 | l,m \in \mathbb{Z}_{2^k-1}, l\neq
  m\}\] has no multiple elements, i.e. all quotients in the field
  representation are pairwise different. Then the orbit of the group
  generated by the companion matrix $P$ of $p(x)$ on $\mathcal{U}$ is
  an orbit code of cardinality $2^{n}-1$ and minimum distance $2k-2$.
\end{lem}
\begin{proof}
  In field representation the elements of the orbit code are:
  \begin{align*}
    C_0 = & \{\alpha^{b_1}, \alpha^{b_2}, \dots, \alpha^{b_{2^k-1}}\}\\
    C_1 = & \{\alpha^{b_1+1}, \alpha^{b_2+1}, \dots, \alpha^{b_{2^k-1}+1}\}\\
    \vdots & \\
    C_{q^n-2} = & \{\alpha^{b_1+q^n-2}, \dots, \alpha^{
      b_{2^k-1}+q^n-2}\}
  \end{align*}
  We show that the intersection between any two code words is at most
  one element. Therefore, assume w.l.o.g. that the first element of $C_h$ is
  equal to the second element of $C_j$:
  \begin{align*}
    \alpha^{b_1+h} = \alpha^{b_2+j} \iff h \equiv b_2-b_1+j \mod
    2^n-1
  \end{align*}
  To have another element in common it has to hold
  \[ b_y +h \equiv b_z +j \mod 2^n-1\] for some $y\neq 1$ (or $z\neq
  2$). Insert $h$ from above:
  \[ b_y +b_2-b_1+j \equiv b_z +j \iff b_2 -b_1 \equiv b_z-b_y \mod
  2^n-1\] By condition the only solution for this equation is $y=1,
  z=2$. Thus there is no second element in the intersection.

  On the other hand one can always find $h\neq j$ such that there is a
  solution to
  \[ b_y +h \equiv b_z +j \mod 2^n-1 , \] hence, the minimum distance is
  exactly $2k-2$.  \qed \end{proof}

As can be found in \cite{el10}, a subspace fulfilling the condition
of Lemma \ref{intone} exists for any $k,n$. Moreover, it is shown how
to combine different orbits to a network code of minimum distance
$2k-2$ and how one-error-correcting decoding can be done.

The result and proof from above can be carried over to arbitrary
finite fields and any starting point $\mathcal{U} \in \G$ in the
following way:
\begin{thm}\label{thmprim}
  Over $\F_q$ let $p(x)$ be a primitive polynomial and $\alpha$ a root
  of it. Assume $\mathcal{U}=\{0,u_1,\dots,u_{q^k-1}\} \in \G$, 
  
  \[\phi(u_i)=\alpha^{b_i} \hspace{0.7cm} \forall i=1,\dots,q^k-1\]
  and $d<k$ be minimal such that any element of the set
  \[\{b_{m}- b_{l} \mod q^n-1 | l,m \in \mathbb{Z}_{q^k-1}, l\neq
  m\}\] has multiplicity less than $q^d-1$, i.e. a quotient of two
  elements in the field representation appears at most $q^d-1$ times
  in the set of all pairwise quotients. Then the orbit of the group
  generated by the companion matrix $P$ of $p(x)$ on $\mathcal{U}$ is
  an orbit code of cardinality $q^{n}-1$ and minimum distance $2k-2d$.
\end{thm}
\begin{proof}
  In analogy to the proof of Lemma \ref{intone}, to have another
  element in common it has to hold
  \[ b_2-b_1 \equiv b_z -b_y \mod q^n-1 .\] By condition there are up to
  $q^d-1$ solutions in $(y, z)$ for this equation, including $y=1,
  z=2$. Thus the intersection of $C_i$ and $C_j$ has at most $q^d-1$
  elements. On the other hand, since $d$ is minimal, one can always
  find $h\neq j$ such that there are $q^d-1$ solutions to
  \[ b_y +h \equiv b_z +j \mod q^n-1 , \] hence,  the minimum distance is
  exactly $2k-2d$.  \qed \end{proof}

If $d=k$, i.e. all quotients are the same, one gets elements with full
intersection which means they are the same element. Then the cardinality of the code is equal to the least element of multiplicity $k$ minus $1$ and the minimum distance has to be computed as before but in the set of all differences that have multiplicity less than $k$.



\subsection{Non-Primitive Generator}


\begin{thm}\label{non1}
  Let $P$ be an irreducible non-primitive companion matrix,
  $\mathfrak{G}$ the group generated by it and denote by $v\mathfrak{G}$ and $\mathcal{U}\mathfrak{G}$ the orbits of $\mathfrak{G}$ on $v\in \F_{q}^{n}$ and $\mathcal{U}\in \G$, respectively.
  If $\mathcal{U} \in \G$ such that 
  \[v\neq w \implies v\mathfrak{G} \neq w\mathfrak{G}  \quad \forall \: v,w \in \mathcal{U} , \]
  then $\mathcal{U}\mathfrak{G}$ on is an orbit code with minimum distance $2k$ and
  cardinality $\mathrm{ord}(P)$.
\end{thm}
\begin{proof}
The cardinality follows from the fact that each element of $\mathcal{U}$ hast its own orbit of
  length $\mathrm{ord}(P)$. Moreover, no code words intersect
  non-trivially, hence the minimum distance is $2k$.  \qed \end{proof}

Note that, if the order of $P$ is equal to $\frac{q^{n}-1}{q^{k}-1}$,
these codes are again spread codes.

\begin{ex}
  Over the binary field let $p(x)=x^4+x^3+x^2+x+1$, $\alpha$ a root of
  $p(x)$ and $P$ its companion matrix. Then $\F_{2^4}\backslash \{0\}$
  is partitioned into
  \[\{\alpha^i | i=0, \dots , 4\} \cup \{\alpha^i(\alpha+1) | i=0,
  \dots , 4\} \cup \{\alpha^i(\alpha^2+1) | i=0, \dots , 4\} .\] Choose
  \begin{align*}
    u_1=&\phi^{-1}(1)=\phi^{-1}(\alpha^0)=(1000)\\
    u_2=&\phi^{-1}(\alpha^3+\alpha^2)=\phi^{-1}(\alpha^2(\alpha+1))=(0011)\\
    u_3=&u_1+u_2=\phi^{-1}(\alpha^3+\alpha^2+1)
    =\phi^{-1}(\alpha^3(\alpha^2+1))=(1011)
  \end{align*}
  such that each $u_i$ is in a different orbit of $\langle P\rangle$
  and $\mathcal{U}=\{0,u_1,u_2,u_3\}$ is a vector space.
  Then the orbit of $\langle P\rangle$ on $\mathcal{U}$ has minimum
  distance $4$ and cardinality $5$, hence it is a spread code.
\end{ex}

\begin{prop}
  Let $P$ and $\mathfrak{G}$ be as before and $\mathcal{U}=\{0,v_{1}, \dots, v_{q^{k}-1}\} \in \G$.
  Let $O_1,...,O_l$ be the different orbits of $\mathfrak{G}$ in $\F_{q}^{n}$. Assume that $m<q^k-1$ elements of $\mathcal{U}$ are in
  the same orbit, say $O_1$, and all other elements are in different
  orbits each, i.e.
 \[v_{i}\mathfrak{G} = v_{j}\mathfrak{G} = O_{1}\quad \forall \: i,j \leq m , \]
  \[v_{i}\neq v_{j} \implies v_{i}\mathfrak{G} \neq v_{j}\mathfrak{G}  \quad \forall \: i,j\geq m . \]
  Apply the theory of Section \ref{prim} to the orbit
  $O_1$ and find $d_1$ fulfilling the conditions of Theorem
  \ref{thmprim}. Then the orbit of $\mathfrak{G}$ on $\mathcal{U}$ is
  a code of length $\mathrm{ord}(P)$ and minimum distance $2k-2d_1$.
\end{prop}
\begin{proof}
  \begin{enumerate}
  \item Since there is at least one orbit $O_i$ that contains exactly
    one element of $\mathcal{U}$, each element of $O_i$ is in exactly
    one code word. Hence the cardinality of the code is
    $\mathrm{ord}(\mathfrak{G})=\mathrm{ord}(P)$.
  \item In analogy to Theorem \ref{non1} the only possible
    intersection is inside $O_1$, which can be found according to the
    theory of cyclic primitive groups.
  \end{enumerate}

  \qed
\end{proof}

We generalize these results to any possible starting point $\in \G$:

\begin{thm}
  Let $P, \mathfrak{G}, \mathcal{U}$ and the orbits $O_1,...,O_l$ be
  as before. Assume that $m_i$ elements of $\mathcal{U}$ are in the
  same orbit $O_i$ ($i=1,\dots, l$). Apply the theory of Section
  \ref{prim} to each orbit $O_i$ and find the corresponding $d_i$ from
  Theorem \ref{thmprim}.  Then the following cases can occur:
  \begin{enumerate}
  \item No intersections of two different orbits coincide. Define $d_{\max}:=
    \max_{i}d_i$.  Then the orbit of $\mathfrak{G}$ on $\mathcal{U}$
    is a code of length $\mathrm{ord}(P)$ and minimum distance
    $2k-2d_{max}$.
  \item Some intersections coincide among some orbits. Then the corresponding $d_i$'s
    add up and the maximum of these is the maximal intersection number $d_{\max}$. 
  \end{enumerate}
Mathematically formulated: Assume $b_{(i,1)},\dots , b_{(i,\mathrm{ord} (P)-1)}$ are the exponents of the field representation of elements of $\mathcal{U}$ on $O_i$. For $i=1,\dots,l$ define 
\[a_{(i,\mu,\lambda)} := b_{(i,\mu)} - b_{(i,\lambda)}\]
and the difference (multi-)sets
\[D_i:=\{a_{(i,\mu,\lambda)}\mid \mu,\lambda \in \{1,\dots, \mathrm{ord}(P)-1\}\},\]
\[D:=\bigcup_{i=1}^l D_i .\]
Denote by $m(a)$ the multiplicity of an element $a$ in $D$ and $d_{\max} := \log_q(\max \{m(a) \mid a\in D\}+1)$.
Then the orbit of $\mathfrak{G}$ on $\mathcal{U}$ is
  a code of length $\mathrm{ord}(P)$ and minimum distance $2k-2d_{\max}$.
\end{thm}


Note that in the case that the minimum distance of the code is $0$ one has
  double elements in the orbit. Then one has to consider the set of different
  code words and compute the cardinality and minimum distance again.

\section{Conclusion}\label{conclusion}


We listed all possible irreducible cyclic orbit codes and showed that
it suffices to investigate the groups generated by companion matrices
of irreducible polynomials. Moreover, polynomials of the same degree
and same order generate codes with the same cardinality and minimum
distance. These two properties of the code depend strongly on the
choice of the starting point in the Grassmannian. We showed how one can
deduce the size and distance of an orbit code for a given subgroup of
$GL_n$ from the starting point $\mathcal{U} \in \G$. For primitive
groups this is quite straight-forward while the non-primitive case is
more difficult.

Subsequently one can use this theory of irreducible cyclic orbit codes to characterize all cyclic orbit codes.

\bibliography{/home/a/rosen/Bib/huge.bib}
\bibliographystyle{plain}

\end{document}